# Statistical Disclosure: Improved, Extended, and Resisted

Navid Emamdoost, Mohammad Sadeq Dousti, Rasool Jalili
Data and Network Security Lab
Department of Computer Engineering, Sharif University of Technology
Tehran, Iran
{emamdoost@ce., dousti@ce., jalili@}sharif.edu

*Abstract*—Traffic analysis is a type of attack on secure communications systems, in which the adversary extracts useful patterns and information from the observed traffic. This paper improves and extends an efficient traffic analysis attack, called "statistical disclosure attack." Moreover, we propose a solution to defend against the improved (and, *a fortiori*, the original) statistical disclosure attack. Our solution delays the attacker considerably, meaning that he should gather significantly more observations to be able to deduce meaningful information from the traffic.

## I. INTRODUCTION

Anonymity is a mechanism for providing privacy in communications systems. Research on anonymous communications systems began by the seminal work of Chaum in 1981. He introduced the concept of *mixes* as a way of providing *unlinkability*, i.e. removing any association between entities in the system [1]. Since then, many anonymity protocols were proposed, each of which has different characteristics for various requirements. Besides designing these protocols, research is conducted on vulnerability analysis of anonymous communications systems. As a result, numerous "privacy-violating" attacks were discovered. The ultimate goal of such attacks is to eliminate or reduce the amount of anonymity. While some attacks exploit the vulnerabilities in the protocol design or implementation—such as timing attacks, tagging attacks, blending attacks, etc. [2], there are other kinds of attacks which assume that the protocol design or implementation has no vulnerabilities. Here, the attacker merely observes protocol execution, while being oblivious to the internal mechanism of anonymity protocol. If the attack is successful, the attacker will be able to find associations between users; that is, he can tell which users where communicating during the attack. These types of attacks are named *Intersection Attacks*. Statistical Disclosure Attack (SDA) is one of the most *efficient* intersection attacks applicable on mix-nets [8].

In this paper, we improve SDA, so as it requires fewer observations to reduce anonymity. For any "target" user Alice, we identify users whose behavior affects Alice's anonymity, and name them *cloak users*. To uncover Alice partners (those associated with her), it is sufficient to merely estimate the behavior of cloak users, rather than all users.

In addition, we extend SDA to cover a non-threshold mix named "SG-Mix". To this end, we use the queueing theory to model message delays in SG-Mix.

We finally suggest a solution to counter both the improved and the original SDA. Our solution delays the adversary considerably, since he needs to gather a great deal of observations before he can effectively mount the attack.

The rest of this paper is organized as follows. In Section II, we present a brief description of mix-nets and traffic analysis attacks, such as disclosure and statistical disclosure attacks. In Section III, we present our contributions: the improved SDA is demonstrated in III.A; III.B includes our extension to the attack to cover a low-latency protocol III.B, and III.C describes our proposed method to resist intersection attacks. We show the effectiveness of the improvement and resistance methods by simulations in Section IV. Finally, Section V concludes the paper.

## II. RELATED WORK

Chaum [3] introduced anonymous communications by defining the concept of "mix." A mix is a special router that provides anonymity by changing the bit pattern of messages, and by reordering them, so that no arriving message can be linked to a leaving one. In the Chaum's mix, the message bit pattern is changed by cryptographic functions, that is, every message is decrypted and will then be sent to its receiver in the alphabetic order. If the mix gets compromised, no anonymity is provided. Hence, instead of using just one mix, usually a chain of mixes is used. The anonymity of messages is guaranteed as long as at least one mix in the chain is honest.

The first model of the mix, introduced in [3], is called "Threshold Mix," as it waits until a constant number of messages arrives, and then decrypts and flushes them out. Other types of mixes, based on their flushing methods include timed mixes, pool mixes, and Stop-and-Go (SG) mixes. A good survey of these flushing algorithms is available in [2]. Here, we review SG-Mix in more detail, as we are going to extend an attack on this protocol.

Kesdogan *et al.*[4] proposed a mix design which does not use batch processing. In this design, the sender derives— from an exponential distribution with parameter μ—a $t_i$ delay for the *i*th mix in the chain, puts this delay in the encrypted message, and sends the message to the SG-Mix. The *i*th SG-Mix decrypts the message and obtains $t_i$. The message will be flushed after a $t_i$ delay. As there is no batch processing, SG-Mix is considered a low-latency mix, and can be used in applications with sensitive timing constraints like web surfing. In [5], Danezis showed that SG-Mix provides the optimal mixing strategy among low-latency anonymity protocols.

Thus far, we have considered mixes as a strategy to provide anonymity in communications networks. Next, we will examine attacks against such anonymous systems.

The ultimate goal of these attacks is to reduce the anonymity of users. Good surveys of such attacks are available in [2][6]. Some attacks exploit vulnerabilities in the protocol design or implementation. The attacker can compromise protocol nodes, trace a message, delay a message or manipulate it to distinguish its relevance to other messages. In addition, the attacker can flood protocol nodes to learn a specific message destination. Moreover, it can replay captured messages from previous executions of the protocol. On the other hand, some other attacks are oblivious to the internal operation of anonymity protocol, and assume there is no vulnerability in the protocol design and implementation. Here, the attacker just observes the protocol execution, and concludes links between users from these observations. The latter type of attack is termed "intersection attacks."

Agrawal and Kesdogan [7] presented an *intersection attack* named "Disclosure Attack." In this attack, the anonymity system is abstracted as follows: a subset of senders sends their messages to a subset of receivers in several rounds. The passive attacker is able to observe messages entering to and leaving from the anonymity system. In each round, two sets of users can be identified: the set of users who send messages (sender-set), and the set of users who receive messages (receiver-set). The attack's aim is to reveal partners of a "target" user, named Alice. It is assumed that Alice has *m* partners. Moreover, she sends at most one message in each round. If Alice is going to send in a round, she selects her receiver uniformly from her *m* partners. The batch size is *b*; that is, in each round, *b* users participate. There are a total of N recipients, and each sender (except Alice) selects its recipient uniformly from all the recipients.

Unfortunately, the attacker in the above model has to do an NP-complete computation to identify Alice partners. As a solution, Danezis [8] proposed a statistical version of the disclosure attack. *Statistical Disclosure Attack* (SDA) is more efficient than the naïve disclosure attack, as it does not have to solve an NP-complete problem. However, SDA cannot identify Alice's partners for certain. There is always some uncertainty in the solution obtained by SDA.

In SDA, users' behavior is modeled by vectors $\vec{u}, \vec{v}$, and $\vec{o}_i$ of N elements. In these vectors, the ith element corresponds to the probability of receiving some specified message by ith user. Vector $\vec{v}$ models Alice's behavior. The ultimate goal of the adversary is to find this vector. In $\vec{v}$, all m partners of Alice have probability $\frac{1}{m}$ and the remaining *N-m* elements are zero. Vector $\vec{u}$ is used to model the behavior of users other than Alice. Initially, all the N elements of this vector are $\frac{1}{N}$. The attacker observes protocol execution and derives observation vectors $\vec{o}_i$ from the receiver-set of round i. The jth element of $\vec{o}_i$ denotes the probability that the jth user be Alice partner in round i. So, in each round that Alice participates, every user in the receiver-set has probability $\frac{1}{b}$ of receiving the message from Alice. Based on the law of large numbers, the attacker can take arithmetic mean of a large number of these observation vectors:

$$\bar{O} = \frac{1}{t}\sum_{i=1}^{t} o_i \approx \frac{\vec{v}+(b-1)\vec{u}}{b}. \quad (1)$$

Accordingly, it can derive $\vec{v}$:

$$\vec{v} \approx \frac{b}{t}\sum_{i=1}^{t} o_i - (b-1)\vec{u}. \quad (2)$$

In [9], authors extended this attack to *Pool Mixes*, which do not flush all messages, and always retain a constant number of messages in the pool. In [10], authors tried to use all observations, and find a better estimation for other users' behavior. They used arithmetic mean of observations in which Alice did not participate, as an alternative to vector $\vec{u}$.

Several methods have also been proposed to counter (i.e. delay) intersection attacks, most of which consider sending dummy messages by users or mixes [10][11][12].

### III. CONTRIBUTIONS

In this section, we express our improvement in SDA's effectiveness, the way to extend it to SG-mixes, and also our proposal to resist against.

The classical SDA has many limiting assumptions; it assumes users (except Alice) can select their partners from all recipients with equal probability. In addition, Alice selects the receiver uniformly from her *m* partners. The mixing process is assumed to be batch processing.

It is more realistic to assume that every user in the anonymity system is associated with a set of partners and the user does not necessarily select the receiver uniformly from this set. Also incorporation of batch processing is a significant merit.

The bottleneck in SDA is the number of observations required to identify Alice partners. So, improving the attack means decreasing the number of such required observations. To do so, we use an estimation to model other users' behavior. SDA was proposed to be applied on anonymity protocols such as mix-nets which use batch processing. The attack was extended [9] to a probabilistic variation of mixes named *Pool Mix*. We extend it to SG-Mix, which is a low-latency anonymity protocol, whilst it delivers the most degree of anonymity among low-latency protocols [5]. Our proposed new method of resisting SDA neither requires the mixing strategy to change, nor requires all users to send dummy messages.

## A. Improving SDA

We propose a more efficient way to estimate the behavior of the other users. To this end, we identify effective users on Alice's anonymity, and then estimate their behavior. In the classical SDA, Alice selects her recipients from the set M, with equal probability, and other users select their recipients uniformly from all *N* recipients. The attacker uses the vector $\vec{u}$ to model the behavior of other users. Therefore, the vector $\vec{v}$, which is the goal of the attack, is an N-ary vector having *m* components of $\frac{1}{m}$ and *N-m* zero components. None of these assumptions are realistic; every user can have her own set of recipients, and is not compelled to select her recipient uniformly from the set.

If we consider a specific set of partners for every user, we cannot use the vector $\vec{u}$ anymore; using $\vec{u}$ implies that users select their recipients uniformly from all the other users. Therefore, estimating the behavior of the other users is necessary. In [10], a solution is proposed to achieve such estimation. They used the *median of observations* from rounds in which Alice has not participated, to model the behavior of other users. The estimation includes all users in the anonymity system; whilst for an effective attack, we should consider only users who are effective in hiding the Alice's communication. Therefore, the method in [10] suffers from hiding of the behavior of users who affect Alice's anonymity. On the other hand, when the number of observations increases, this estimation becomes very similar to $\vec{u}$, and so is not an appropriate model.

In the rest of this paper, we use the notations in Table 1 to illustrate our method.

TABLE I. SYMBOLS USED IN THIS PAPER

| Symbol | Descriptions |
|---|---|
| A | Set of all senders |
| $A'_i$ | Sender-set in round i |
| $G_i$ | Set of rounds in which user i participated |
| B | Set of all receivers |
| $B'_i$ | Receiver-set in round i |
| $\vec{v}$ | Alice's send vector |
| $\vec{u}$ | Uniform send vector |
| $\vec{o}_i$ | Observations vector in round i |
| $o_{i,j}$ | jth element of ith observation vector |
| $a_r$ | Number of Alice's messages in round r |

To be more precise, the following definitions are considered.

*Definition 1:* S is a Boolean function demonstrating if the user *e* has participated in round *i*:

$$S(e,i) = \begin{cases} 1 & \text{if user } e \text{ has participated in round } i \\ 0 & \text{otherwise.} \end{cases}$$

Using *S*, the sender-set of the *i*th round can be defined as follows.

*Definition 2:* $A'_i$ is the set of users participating in round *i*:

$$A'_i = \{e \in A \mid S(e,i) = 1\}$$

Supposing the protocol to be executed for *t* rounds, we can define the set $G_i$ as follows.

*Definition 3:* $G_i$ is the set of all rounds in which user $e_i$ has participated:

$$G_i = \{j \mid 1 \leq j \leq t \wedge e_i \in A'_j\}$$

From Definition 3, we can divide rounds into two categories: those including Alice ($G_{Alice}$) and those excluding Alice ($\bar{G}_{Alice}$). We need an estimate on the behavior of users who affect Alice's anonymity.

When Alice participates in a round, her messages become anonymous among the other messages sent in that round. So, all users participating in the round affect the anonymity of Alice's messages. From the attacker point of view, such effective users must be identified and be considered in disclosing Alice partners.

In each round, attacker notices a set of senders who send their messages to a set of receivers. So, the sender-set and the receiver-set can be constructed for the round. If Alice participates in the *i*th round, every user in the *i*th sender-set affects the Alice's anonymity. To identify such users, we define cloak users as follow:

*Definition 4:* for the user $e_i$, the set of cloak users include users who have participated in at least one round with $e_i$.

$$Cloak_i = \{e_j \mid e_j \in \bigcup_{k \in G_i} A'_k - \{e_i\}\}$$

The set of Alice's cloak users is $Cloak_{Alice}$. We need an estimate on the behavior of users in $Cloak_{Alice}$. To this end, we extract such rounds from $\bar{G}_{Alice}$ in which at least one user from $Cloak_{Alice}$ has participated. Such rounds are collected in the set $P_{Alice}$.

*Definition 5:* The set $P_i$ includes those rounds of $\bar{G}_i$ in which the cloak user of $e_i$ has participated:

$$P_i = \{j \in \bar{G}_i \mid Cloak_i \cap A'_j \neq \varnothing\}$$

The average of the observations from rounds in $P_{Alice}$ is used as the estimation for the behavior of cloak users ($\overrightarrow{CloackUsr}$) calculated as:

$$\overrightarrow{CloakUsr} = \frac{1}{|P_{Alice}|} \sum_{i \in P_{Alice}} \vec{o}_i . \quad (3)$$

Using this estimation and the *law of large numbers*, the vector $\vec{v}$ can be calculated as:

$$\vec{v} = \frac{b}{a t'} \sum_{i \in G_{Alice}} \vec{o}_i - \frac{(b-\bar{a})}{\bar{a}} \overrightarrow{CloakUsr} . \quad (4)$$

Where $t' = |G_{Alice}|$ and $\bar{a}$ is the average of Alice's share in the number of messages sent in each round:

$$\bar{a} = \frac{1}{t'} \sum_{i \in G_{Alice}} a_i . \quad (5)$$

The improved SDA can be summarized as the following steps:
1. Observe all rounds of the protocol execution
2. Calculate observation vectors for each round
3. Divide rounds into two sets: $G_{Alice}$ and $\bar{G}_{Alice}$

4. Identify cloak users
5. Construct $P_{Alice}$
6. Calculate $\overrightarrow{CloackUsr}$
7. Calculate $\vec{v}$

To this end, the attacker uses all available observations to identify Alice partners. Using this improved attack, a better estimation of behavior of effective users on Alice's anonymity will be achieved; and no constraint will be imposed on any other users' behavior. In Section IV, we explain how this method reduces the number of required observations.

### B. Extending SDA

SDA has been applied to anonymity systems with batch processing. In [9], authors extended SDA to cover Pool Mixes, which is a probabilistic mix. In this section, we extend the attack to cover a low-latency anonymity protocol named SG-Mix [4]. We chose SG-Mix due to its provision of the most optimal mixing strategies among low-latency protocols [5].

The model of anonymity system in SDA is a simplified form of mix-net, where execution of protocol is divided into rounds. In each round, a subset of users sends their messages to another subset of users. So, in protocols such as mix-nets, which collect input messages into batches and processes them after a threshold, each batch is equivalent to a round. However, batch processing imposes a great delay to the delivered messages. As most of online applications have timing constraints, low latency anonymity protocols would be of more interest.

For simplicity, we extend the attack to one SG-Mix, and assume whenever Alice sends a message to SG-Mix, she will not send another one until the first one has left the SG-Mix ($a_r=1$).

To apply SDA to any anonymity protocol, the protocol must be mapped onto the attack model. To extend the attack to SG-Mix, it is necessary to identify sender-sets and receiver-sets of each round. To do so, we model the delay of a SG-Mix.

#### 1) SG-Mix Delay Model

As mentioned in Section II, the distribution of message delays is exponential with parameter μ. Assuming the message arrival distribution is Poisson with parameter λ, then the SG-Mix can be modeled as an M/M/∞ queue. The maximum delay of a message (we refer to it as τ) can be found using such a modeling.

As distribution of delay in SG-Mix is exponential with parameter μ, the mean and standard deviation of the delay is $\frac{1}{\mu}$. Using one-sided Chebychev's inequality, the maximum delay of a message in SG-Mix can be estimated. The inequality says:

$$\Pr(X + E(X) \geq k\sigma) \leq \frac{1}{1+k^2}. \qquad (6)$$

where $X$ is a random variable with mean $E(X)$ and standard deviation $\sigma$. The parameter $k$ determines the amount of confidence. For example, for $k=3$, at least 90% values of $X$ are less than or equal to $E(X)+k\sigma$. In such a case, with the confidence of $1 - \frac{1}{1+k^2}$, it can be said that a message will leave SG-Mix at most $\frac{k+1}{\mu}$ after its entrance. In the other words:

$$\tau = \frac{k+1}{\mu}. \qquad (7)$$

To find τ, μ is required; which is the parameter of delay distribution. In an M/M/∞ queue, the number of messages in the queue, upon entrance of a new message is $\frac{\lambda}{\mu}$. To provide anonymity for messages, it is necessary that at least one message be in the queue upon its entrance; i.e. $\frac{\lambda}{\mu} \geq 1$. Otherwise, SG-Mix behaves as a first-in first-out queue and no anonymity will be provided [5], i.e:

$$\mu \leq \lambda. \qquad (8)$$

The parameter λ can be anticipated from the rate of incoming messages. So, λ is used as an upper bound for μ in our relations. To estimate λ, incoming messages to SG-Mix can be observed for a period of time *T*. Considering the number of entered messages in this period as $x$, $\lambda = \frac{x}{T}$; having λ as the message entrance rate in the unit of time.

#### 2) Construct Sets

Using the maximum delay from (7), we can build sender-set and receiver-set. The set of all messages to which Alice's message blend forms receiver-set; and the set of all potential senders of messages in receiver-set, forms sender-set. The following theorems illustrate formation of these two sets.

*Theorem 1:* If Alice's message enters SG-Mix at time *t*, the receiver-set is comprised of all users receiving at least one message in the interval $W_1 = [t, t + \tau]$.

*Proof:* As the maximum delay of a message in SG-Mix is τ, Alice's message will leave the Mix at most at time $t + \tau$. So, every user receiving at least one message from Mix in $[t, t + \tau]$ interval, can be Alice's partner, thus must be in the receiver-set. □

*Theorem 2:* If Alice's message enters SG-Mix at time *t*, all users who send a message to the Mix in the interval $W_2 = [t - \tau, t + \tau]$ will be in the sender-set.

*Proof:* From Theorem 1, the receiver-set is formed. So we try to relate messages in the receiver-set to senders. As the maximum delay is τ, so the message leaving the Mix at time *t*, could have been sent at worst at $t - \tau$ (with maximum delay). The message which leaves Mix at time $t + \tau$, could have been sent at $t + \tau$ (with no delay). Thus, every user who sends a message to Mix in the interval $[t - \tau, t + \tau]$ interval, is a member of the sender-set. □

To perform the attack, observation vectors ($\vec{o}$) are necessary. As mentioned earlier in Section II, these vectors can be calculated from the receiver-sets. Cloak users are identified from sender-sets. The only remaining parameter prior to performing the attack is the batch size (*b*). In SDA, this parameter indicates the number of sent messages in each batch; but in SG-Mix there is no batch processing. The batch size determines number of messages which are blended together to provide anonymity. The size of the sender-sets can be used as parameter *b*. The number of users participating in each sender-set differs from round to round; and so we use the mean size of sender-sets as parameter $b$.

The extension of SDA to SG-Mixes can be summed up through the following steps:
1. Observe the protocol execution
2. Estimate the parameter $\lambda$
3. Calculate the intervals $W_1$ and $W_2$
4. Determine the sender-set and the receiver-set for Alice's messages
5. Obtain parameter $b$ from the size of sender-sets
6. Calculate the vectors $\vec{o}$ from receiver-sets
7. Identify cloak users from sender-sets
8. Determine the sender and receiver-sets for cloak users
9. Calculate $\vec{o}$ vectors for cloak users
10. Calculate $\overrightarrow{CloakUsr}$
11. Calculate $\vec{v}$

### C. Resisting Against the Improved SDA

There is no way to fully defend a protocol against intersection attacks such as SDA; whereas some ideas have been proposed to delay the attack; that is, increasing the number of required observations to succeed. Almost all of such methods include sending *dummy messages* by mix or by users in each round. However, the methods suffer from imposing of a great communication load due to sending of dummy messages. In some methods, users must participate in each round despite their willingness. The mix strategy must also be changed (to send/drop dummy messages) and thus the mixing process takes longer.

As mentioned in Section IV, increasing the value of two parameters $b$ and $m$ leads to an increase in the number of required observations. Sending dummy messages by users or mix tends to benefit from the role of parameter $b$. This parameter determines the number of messages blending together in each round. So, sending dummy messages is an attempt to hide Alice's message among more cloak users. Our Sybil defense benefits from the role of parameter $m$ in the attack difficulty. The parameter determines the diversity of Alice's selection. More divert selections leaks less information about each recipient's identity.

In our method, Alice participates with $n$ pseudonyms in the anonymity protocol. These pseudonyms are not linkable to Alice's identity in any way. Every pseudonym has its own set of recipients (different $m$). In each round where Alice intends to send a message, she sends a message by any of her pseudonym as well. For example, if Alice participates in the protocol as her own identity as well as another identity like $A'$; whenever Alice wants to send a message, she selects her recipient from a set of $m$ members, and $A'$ selects her recipient from a set of $m'$ members. In each round where Alice is included in the sender-set, $A'$ is also included; accordingly the attacker cannot distinguish partners of $A'$ from partners of Alice. This can be considered as a scenario in which a user selects her partners from a set of size $m + m'$.

More pseudonymous identities lead to more complexity for the attacker. We refer to our method as Sybil defense, because of using multiple and unlinkable identities.

In Section IV, we will describe the implementation and evaluation of our defense method against the standard and improved SDA. As advantages of our defense method, there is no need to overwhelm the network by dummy messages and also, each user can use such protection by her willing and based on the amount of required anonymity.

## IV. SIMULATION

Our simulator is inspired from [10]. Similar simulators was also used in [12] and [13].

### A. Simulation Plan

For simulation, we generated observations and applied the standard and improved SDA to a unique system *configuration* under assumptions of original SDA. By configuration, we mean specific values of system parameters $N$, $b$, and $m$. To create observations for a configuration, we determined the set of partners for all users. To simulate each round, a sample of size $b$ was obtained with replacement from the set of all users. This sample denotes the set of senders in that round. The sampling was performed *with replacement* to allow each user to send more than one message per round. For each sender, the corresponding receiver(s) was (were) determined based on the sender's partner set.

To show effectiveness of our improvement, we applied both attacks to several configurations and compared the number of required observation for each successful attack. We continued an attack until it revealed at least 80% of Alice's partners or passed over the limit of 5000 observations. We repeated the experiment 100 times per configuration before reporting the attack results. The pseudo code of the simulation is shown in Figure 1.

### B. Results

We run the simulation with default values of $N=20000$, $b=50$, and $m=20$. In each comparison, we changed one parameter, whereas the other two were kept unchanged.

Figure 2 depicts the effect of changing $N$ (number of system users) on the attacks. As $N$ increases, the number of required observations decreases for both the attacks; with the need of fewer observations in the case of the improved SDA.

Figure 3 demonstrates the effect of batch size on the attacks. As parameter $b$ increases, the number of required observations increases. As increasing the batch size results in increasing the number of messages which blend with Alice's ones, the problem become more difficult. As batch size increases, the improved attack needs less observation than the standard SDA.

Figure 4 shows the role of parameter $m$ on the attacks. This parameter defines the number of Alice partners. By increasing $m$, Alice sends messages to more different users, so Alice's behavior becomes more diverse. For SDA, the number of observations increases extraordinarily when $m \geq 20$, but for the improved SDA, the number of observations increases gently. In the other words, when $m \geq 20$, SDA becomes inapplicable; while the improved SDA is still applicable. Accordingly, the main advantage of our improvement is where Alice increases her behavior diversity.

```
#Generate Observations
for i = 1 .. t
        senderSet← a sample of size b from N
        for each k in senderSet do
                Choose a receiver from k's send vector
                and insert it into receiverSet
        end
end

# Statistical Disclosure Attack
rounds← rounds that Alice participated in it
t ← |rounds|
for each r in rounds
        for each k in receiverSet(r)
                O(k) += 1/b
        end
end
```
$$\bar{O} \leftarrow \sum_k \frac{O(k)}{t}$$
$$\vec{v} \leftarrow b.\bar{O} - (b-1).\vec{u}$$

```
# Improved SDA
Calculate O vectors for all rounds, as before.
for each r in rounds
        if Alice is a sender in r
                then G ← G∪r
                else Ḡ ← Ḡ∪r
        end
for each r in Ḡ
        add all senders of r except Alice to Cloak_Alice
end
for each r in Ḡ
        if Cloak_Alice ∩ A'_r ≠ ∅
                then P ← P ∪ r
end
```
$$\overrightarrow{CloakUsr} \leftarrow \frac{1}{|P|} \sum_k (O(k)|k \in P)$$
$$\vec{v} \leftarrow \frac{b}{|G|} \sum_k (O(k)|k \in G) - (b-1)\overrightarrow{CloakUsr}$$

Figure 1. Simulation pseudo code

To show the effect of the Sybil Defense introduced in Section III.C, we changed the way of generating of observations. In all the related simulations, we used just one pseudonym ($A'$) for Alice. To determine the senders of a round, if Alice is included as a sender, we inserted the user $A'$ to the set of senders and for every user in the sender-set, the corresponding recipient is selected from the partner set. The simulations confirmed that no attack could finish in none of the configurations before reaching the observation limit. This means that the attacker is unable to identify at least 80% of Alice's partners, whereas prior to employing the Sybil defense, both attacks were finished before reaching the 5000 observation limit. So the Sybil defense can effectively delay the attacker's success.

While we examined our resistance method on SDA, it is effective on all intersection attacks, such as the Hitting Set Attack [14], as it can unsettle the uniqueness of hitting set.

## V. CONCLUSION AND FUTURE WORKS

In this paper, we improved the Statistical Disclosure Attack (SDA) in a way that it needs fewer observations to succeed, and extended it to cover a non-threshold mix protocol called SG-Mix. Finally, we proposed a resistance method to delay attackers' success.

We relaxed some limiting and unrealistic assumptions of SDA. In the relaxed assumptions, other users can have their specific sending vectors, and are not compelled to select their receivers uniformly from all users. Moreover, Alice can select her receivers non-uniformly from her m partners.

To model the message delay in SG-Mix we employed the queueing theory. An SG-Mix can be modeled as an M/M/∞ queue with Poisson arrivals and exponential delays. From there the maximum delay of a message was calculated and then observation vectors were constructed.

Our resistance method resulted in a great effect on the attack. In the normal scenario, the attacker was able to identify at least 80% of Alice's partners, while by employing just one pseudonymous identity by Alice, the attack became inapplicable. We believe that our method is effective on all intersection attacks such as the hitting set attack (by unsettling the uniqueness of hitting set), but this must be evaluated in more details as future work.

[12] N. Mallesh, and M. Wright, "Countering Statistical Disclosure with Receiver-Bound Cover Traffic," in *Computer Security – ESORICS 2007*. vol. 4734, J. Biskup and J. López, Eds., ed: Springer Berlin / Heidelberg, 2008, pp. 547-562.

[13] D. Kedogan, D. Agrawal, and S. Penz, "Limits of anonymity in open environments," in *Proceedings of Information Hiding Workshop (IH 2002)*, 2002, pp. 53-69.

[14] D. Kesdogan, and L. Pimenidis, "The hitting set attack on anonymity protocols," in *Proceedings of 6th Information Hiding Workshop (IH 2004)*, 2004, pp. 326-339.

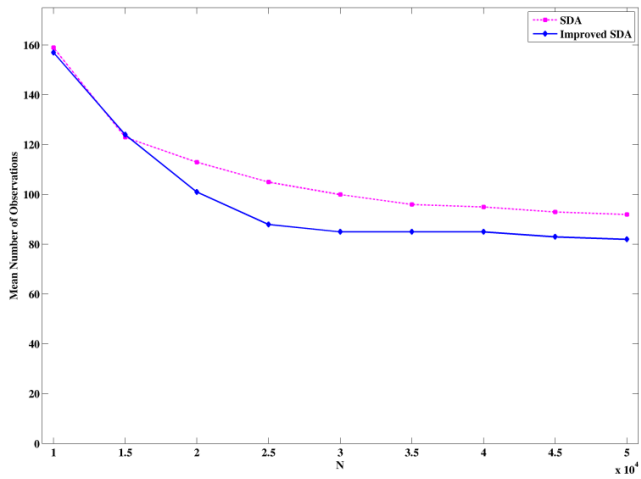

Figure 2. Effect of parameter N (number of users) on attacks

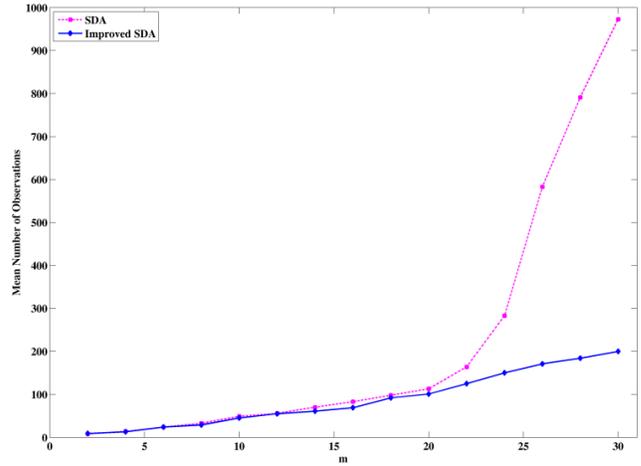

Figure 4. Effect of parameter m (number of partners) on attacks

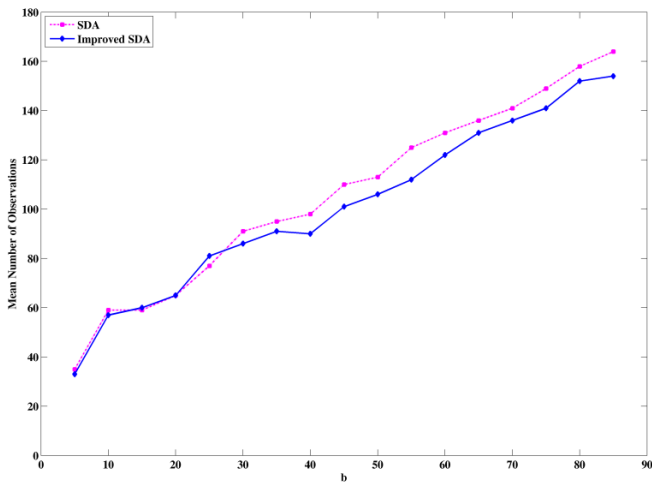

Figure 3. Effect of parameter b (batch size) on attacks